\documentclass[a4paper,11pt]{article}
\pdfoutput=1 

\usepackage{jcappub} 
\usepackage[T1]{fontenc} 
\usepackage[utf8]{inputenc}
\usepackage{hyperref}       
\usepackage{url}            
\usepackage{booktabs}       
\usepackage{amsfonts}       
\usepackage{nicefrac}       
\usepackage{microtype}      
\usepackage{amsmath}
\usepackage{tikz}
\usepackage{listofitems}
\usepackage{xcolor}
\usepackage{graphicx}
\usepackage{wrapfig}
\usepackage{graphicx}
\usepackage{amssymb,amsmath, eufrak}
\usepackage{lineno}
\usepackage{xcolor,bm}

\usepackage[]{color}
\definecolor{Red}{rgb}{0.9,0.17,0.31}
\definecolor{Green}{rgb}{0,0.5,0}
\definecolor{ashgrey}{rgb}{0.7, 0.75, 0.71}
\definecolor{bluebell}{rgb}{0.64, 0.64, 0.82}
\definecolor{darkorchid}{rgb}{0.6, 0.2, 0.8}
\definecolor{glaucous}{rgb}{0.38, 0.51, 0.71}
\definecolor{darkcerulean}{rgb}{0.03, 0.27, 0.49}
\definecolor{flame}{rgb}{0.89, 0.35, 0.13}
\definecolor{brightmaroon}{rgb}{0.76, 0.13, 0.28}
\definecolor{maroon}{rgb}{0.5, 0.0, 0.0}
\definecolor{carmine}{rgb}{0.59, 0.0, 0.09}
\definecolor{darkcyan}{rgb}{0.0, 0.55, 0.55}
\definecolor{electricultramarine}{rgb}{0.25, 0.0, 1.0}
\definecolor{mediumturquoise}{rgb}{0.28, 0.82, 0.8}
\definecolor{lightseagreen}{rgb}{0.13, 0.7, 0.67}

\usepackage{subfigure}
\usepackage{lineno}
\usepackage{soul}
\usepackage{bm}
\usepackage{hyphenat}
\usepackage{seqsplit}
\usepackage{graphicx}
\usepackage{amsfonts}
\usepackage{amssymb,amscd}

\usepackage{xcolor}
\usepackage{comment}
\usepackage{amsmath}
\usepackage{dcolumn}
\usepackage{bm}

\title{\boldmath Nonparametric signal separation in very-high-energy gamma ray observations with probabilistic neural networks}

\author[1]{M. Ullmo,}
\author[1]{E. Moulin}
\affiliation[1]{Irfu, CEA Saclay, Universit\'e Paris-Saclay F-91191 Gif-Sur-Yvette Cedex, France}
\emailAdd{marion.ullmo@gmail.com}
\emailAdd{emmanuel.moulin@cea.fr}

\abstract{
An intriguing challenge in observational astronomy is the separation signals in areas where multiple signals intersect. A typical instance of this in very-high-energy (VHE,  E$\gtrsim$100 GeV) gamma-ray astronomy is the issue of residual background in observations. This background arises when cosmic-ray protons are mistakenly identified as gamma-rays from sources of interest, thereby blending with signals from astrophysical sources of interest.
We introduce a deep ensemble approach to determine a non-parametric estimation of source and background signals in VHE gamma observations, as well as a likelihood-derived epistemic uncertainty on these estimations.
We rely on minimal assumptions, exploiting the separability of space and energy components in the signals, and defining a small region in coordinate space where the source signal is assumed to be negligible compared to background signal.
The model is applied both on mock observations, including a simple toy case and a realistic simulation of dark matter annihilation in the Galactic center, as well as true observations from the public H.E.S.S. data release, specifically datasets of the Crab nebula and the pulsar wind nebula MSH 15-52.
Our method performs well in mock cases, where the ground truth is known, and compares favorably against conventional physical analysis approaches when applied to true observations.
In the case of the mock dark matter signal in the Galactic center, our work opens new avenues for component separation in this complex region of the VHE sky.
}

\begin{document}
\maketitle
\flushbottom

\section{Introduction}\label{sec:intro}
Very-high-energy (VHE, E$\gtrsim$100 GeV) gamma rays are a versatile tool for probing the most violent phenomena in the Universe. They can reveal information about the regions they traverse and the mechanisms that produce them, such as the acceleration of relativistic particles through plasma shocks and magnetic fields, or processes like dark matter annihilation.
Identifying and characterizing such mechanisms requires a careful analysis of the gamma ray sources, notably through the examination of their spatial, spectral, and temporal features, prompting the need for precise analytical tools to accurately capture them.
However, at these very high energies the Earth's atmosphere is fully opaque. Thus ground-based telescopes must observe VHE gamma rays indirectly, one event at a time, by detecting the Cherenkov light emitted by the extensive air shower produced when a ray interacts with the atmosphere. These Imaging Atmospheric Cherenkov Telescopes (IACTs) include instruments like the High Energy Stereoscopic System (H.E.S.S.~\cite{HESS}), Major Atmospheric Gamma Imaging Cherenkov (MAGIC~\cite{MAGIC}) telescopes, the Very Energetic Radiation Imaging Telescope Array System (VERITAS~\cite{VERITAS}), and the upcoming Cherenkov Telescope Array (CTA~\cite{cta2018science}), which promises to revolutionize the field with unprecedented sensitivity and resolution.
\begin{figure}[htbp]
    \centering
    \includegraphics[width=\textwidth]{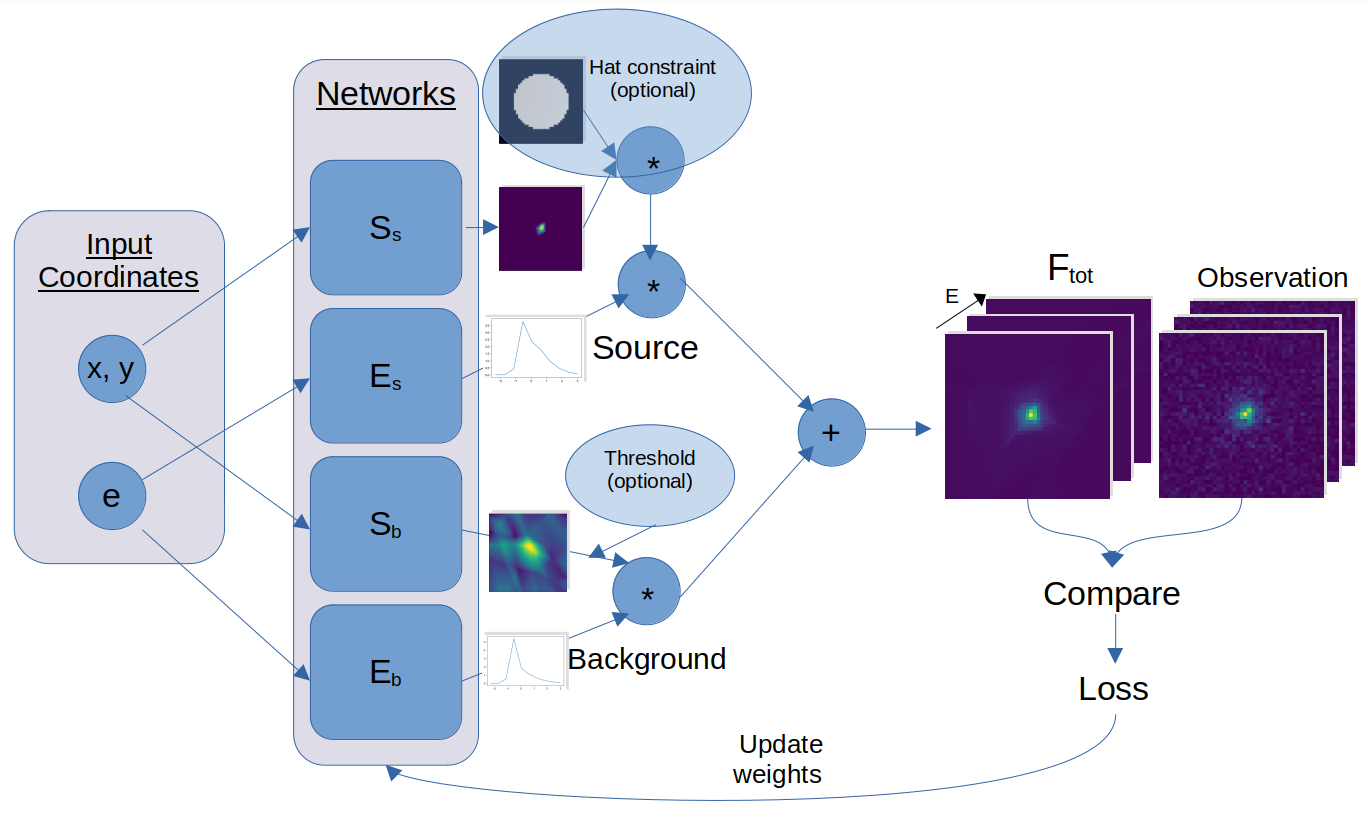}
    \caption{The training setup. The networks output an estimation of $S_s, E_s, S_b$ and $E_b$ over the region of interest in the coordinate space. These are combined to make $F_{tot}$, which is then compared to the observation through the loss given in Eq.~(\ref{eq:Loss_developped}). Optionally, constraints are imposed on $S_s$ and $S_b$.}
    \label{fig:NN_setup}
\end{figure}

A critical challenge in ground-based gamma-ray astronomy is the contamination from high-energy cosmic rays (CR), coming from all directions of the sky. These CR create showers of secondary particles in the Earth's atmosphere that produce Cherenkov radiation similar to that of gamma-ray induced showers. As a result, observations of gamma-ray signals are often contaminated by this background noise.
Mitigating this background noise can be approached in two distinct ways: the first involves a classification task performed during event reconstruction, utilizing the shower images captured by the telescopes to separate gamma showers from hadronic ones. The second approach, typically complementary to the first, involves a statistical estimation of the residual background in control regions using the reconstructed event list after the reconstruction process.

In the classification approach, several techniques have been developed, ranging from standard analytical methods like the Hillas parameters method~\cite{hillas2013evolution, weekes1989observation} to machine-learning techniques like random forest algorithms~\cite{Albert:2007yd,bock2004methods}, boosted decision trees~\cite{ohm2009gamma,fiasson2010optimization,krause2017improved, voss2009tmva} or graph networks~\cite{glombitza2023application}. While these methods prove highly effective and considerably reduce the amount of CR misidentified as gamma rays, none are as of yet $100\%$ effective. Consequently, even processed observations still contain a residual background, which is spatially isotropic and often stronger than the sources of interest.

To address the issue of residual background in the event list, several statistical methods have been developed, depending on whether the source position is known \textit{a priori}. 
One common technique is the \textit{Ring Background} method~\cite{Berge:2006ae}, where a ring-shaped region around the \textit{on}-source area (region of the observation containing the source signal) is used to estimate the background. This approach assumes that the background is isotropic and can be accurately represented by the measurements in the surrounding region provided all VHE emissions there are properly excluded for the background estimate within the on region. \textit{Adaptative Ring background} methods have been developed when exclusion regions cover a significant part the field of view~\cite{HESS:2018pbp}.
Another method is the multiple OFF technique~\cite{Berge:2006ae}, which uses multiple OFF-source regions to obtain a more reliable background estimate by averaging out spatial fluctuations. Advanced statistical methods, such as the \textit{template-background method}~\cite{2003A&A...410..389R}, involve creating detailed models of the background based on the distribution of cosmic-ray events across the field of view.
Despite these sophisticated techniques, limitations persist. The assumption of isotropy in the ring background method may not always hold true, particularly in regions with complex background structures. The multiple OFF technique, while reducing statistical fluctuations, requires careful selection of OFF-source regions to avoid biases. The template background method demands extensive computational resources and accurate modeling, which can be challenging given the variability of cosmic-ray events.
Consequently, while these methods significantly mitigate background contamination, they do not completely eliminate it, prompting the need to seek alternative solutions.

Neural networks (NNs~\cite{schmidhuber2015deep}) have demonstrated their effectiveness in such signal separation tasks across diverse fields, including acoustics~\cite{wang2018supervised}, seismic monitoring~\cite{zhu2019seismic}, medical imaging~\cite{xu2019deep}, and even astronomy~\cite{wang2022galaxy,burke2019deblending}.

In this article we introduce a proof of concept for a NN-based framework to address the issue of CR background contamination, and as a first step for the more general problem of multiple component separation in VHE astrophysics.
Our approach consists in jointly estimating the mean fields of the signals corresponding to CR background and source of interest to maximize the likelihood of a given observation, while providing a statistically-meaningful uncertainty to account for model limitations as well as statistical noise in the observations.
Our method offers several significant advantages. It avoids making any prior parametric assumptions about component shapes, making it ideal for sources with complex shapes, and does not require a training database, relying solely on the observation of interest for signal and background estimation; conversely, this means that training is required for each individual case, contrary to common machine learning approaches. Additionally, it is flexible and can easily incorporate constraints based on prior knowledge to improve the estimation process.
The paper is organized as follows: Section~\ref{sec:nn} addresses the problem setup and the description of our neural network model.
Section~\ref{sec:dataset} presents the data on which the model is applied.
The results are presented in section~\ref{sec:results}. In section~\ref{sec:conclusion} we further interpret results, address limitations and discuss outlooks of the present work.

\section{Method and Neural Network framework}
\label{sec:nn}

\subsection{Problem setup and formalism}
We consider a given observation, a 3D histogram of events $(e,\vec{x})$ binned in energy and space, as a Poisson realization of an underlying mean field $F_{tot}$.
We assume $F_{tot}$ is made up of two components, one for source emission of interest ($F_s$) and the other due to the background noise ($F_b$).
We additionally assume that both $F_s$ and $F_b$ can be further decomposed into space and energy components $(S_s,E_s)$ and $(S_b,E_b)$, respectively, such that:
\begin{equation}
  \begin{cases}
F_s(\vec{x},e)=S_s(\vec{x})\times E_s(e),\\
F_b(\vec{x},e)=S_b(\vec{x})\times E_b(e)\\
  \end{cases}
\end{equation}
This gives the signal mixture expressed as :
\begin{equation}\label{eq:Ftot}
    F_{tot}(\vec{x},e)=S_s(\vec{x})\times E_s(e)+S_b(\vec{x})\times E_b(e)\, .
\end{equation}
As a convention, we impose that $\sum_{obs}E_{s,b}=1$, such that $E_{s,b}$ are dimensionless and physical quantities are expressed within $S_{s,b}$. 

\begin{figure}[h!]
    \centering
    \begin{tabular}{c|c|c}
    \hline
    \hline
    Networks & $S_s,S_b$ & $E_s,E_b$\\
    \hline     \hline
    Input dimension and coordinate & 2, $\vec{x}$ & 1, $e$\\
    \hline
    Number of layers & $5 (S_s), 4 (S_b)$  & $4$ \\
    \hline
    Neurons per layer & $10$ & $5$ \\
    \hline
    Layer Activation function& Leaky ReLU & Leaky ReLU \\
    \hline
    Final Activation function& ELU & ELU \\
    \hline
    Output dimension & 1 & 1 \\
    \hline
    \hline
    \end{tabular}
 \caption{Hyperparameter settings in the network architecture used in this work. All Neural Networks are trained using the {\it Adam} optimizer~\cite{2014arXiv1412.6980K}  with a learning rate $lr$=10$^{-3}$.}
 \label{tab:archis}
\end{figure}

We aim to estimate the spatial and spectral shapes $S_s, E_s, S_b$ and $E_b$ from the 2-component mixture in our observed field of view.
To do so, we build four distinct NNs to emulate the four components.
The networks are simple fully connected (dense) neural networks (see architectures in table \ref{tab:archis}).
They take input coordinates ($\vec{x}$ for $S_s$ and $S_b$, $e$ for $E_s$ and $E_b$), and their outputs (estimates of $S_s, E_s, S_b$ and $E_b$ over the observed region) are combined as in Eq.~(\ref{eq:Ftot}) to form $F_{tot}^{est}$, which is then compared to the observation. All network weights are updated simultaneously to minimize the loss described below.

Given that we aim to reconstruct $F_{tot}$ such that our observation is maximally probable with respect to $F_{tot}$, our loss, which we aim to minimize, is thus chosen to be:
\begin{equation}\label{eq:Loss_simple}
    L=-\langle\log{\mathcal{P}(Obs|F_{tot})}\rangle_{\rm bins}
\end{equation}
Considering  probabilities over individual bins of coordinates $(\vec{x},e)$ this becomes:
\begin{equation}\label{eq:Loss_developped}
    L=-\frac{1}{n_{\rm bins}}\sum_{\vec{x},e}\log{\mathcal{P}(Obs(\vec{x},e)|F_{tot}(\vec{x},e))} \, ,
\end{equation}
where $\mathcal{P}(n|m)=\frac{e^{-m} m^n}{n!}$ is the Poisson probability of observing $n$ counts given an average occurrence $m$, consistently with the expected statistics of an observation.

The network architectures are deliberately designed to be relatively shallow and low in parameters.
This low parametrization imposes a certain smoothness on the output functions and helps prevent overfitting, which would occur if the network were complex enough to fit the Poisson noise in the observations.
We choose to use a separate dense net for each component for better control of this parametrization, as well as to keep them independent by default.
We anticipate the background to be smoother in space compared to the source, justifying the use of a lower-parameter architecture. Future research should explore different architectures, including using a single combined net with four outputs rather than four independent ones, weight initialization methods, learning rates and gradient descent algorithms, as the current choices have been made empirically and may benefit from optimization.
For full implementation details, the code is available on GitHub~\cite{gitmarion}.

\subsection{Network Training}
\subsubsection{Deep ensemble approach}
Our approach essentially equates to a nonparametric fitting of source and background, with a solution that maximizes likelihood.
However, given the high dimension of our problem we can expect a few complications.
First, whether parameter space has a single global maximum; given the redundancy in the NNs we can already expect a manifold of parameter configurations that output the same functions/predictions, however depending on several factors (noise in the observation, overlapping functions, NN constraints, etc), we can additionally expect a manifold of equally-likely solutions that correspond to the global maxima.
Additionally, depending on initial conditions we can expect networks to find and get stuck into local maxima during training.

To account for local maxima cases and to better capture the manifold of equiprobable maxima, we opt for a deep ensemble approach; instead of training a single model (made up of four NNs), we train a set of models and observe their combined results.
In our case, we train 10 separate models. For each we only change initial weights, which are randomized from a Gaussian distribution. 
Depending on the training data, the models are trained for $10^4$ to $10^5$ gradient updates, taking approximately 2 to 20 minutes per model. These timings were measured on a local machine with an AMD Ryzen 5 5600H CPU.
A more comprehensive approach would be to vary NN architectures for each separate model as well, and experiment with varying set size; it is let for further study.

To get a final estimate for all functions, we compute a weighted average of all estimations. We additionally obtain a model uncertainty for all functions by computing a weighted standard deviation. The weights for each estimation $w_i$ are defined relative to estimation likelihood in the following way:
\begin{equation}
    w_i=\exp{\Bigg(\frac{1}{c_{\rm tot}}\sum_b c_b \log{\mathcal{P}\Big(obs(b)|F_{tot}^i(b)\Big)}\Bigg)}\, ,
\end{equation}\label{eq:ponderation}
where $c_{\rm tot}$ is the total particle count, $c_b$ is the particle count in bin $b$, and $F_{\rm tot}^i(b)$ is the total estimated field at bin $b$ for estimation $i$.
In other words, every weight $w_i$ is itself a weighted log-average probability per bin of the observation with respect to $F_{\rm tot}$, where every bin is weighted by its particle count, such that regions of the coordinate space with higher statistics are conferred more weight. 

\subsubsection{Adding constraints}
Additionally we need to impose specific constraints on the model to differentiate source and background signal (i.e., to break the symmetry in the model as made explicit by Eq.~(\ref{eq:Ftot})) and to leverage all available knowledge for more precise estimations. We opt in this work for the following two constraints. 

\paragraph{Hat constraint}
A first simple way to do this is to define an OFF region, where source signal is expected to be negligible relative to background. Having defined this region, we can impose our source signal to be zero within it simply by multiplying its space component with a "hat" function $H$ such that $H(\vec{x})=0$ in OFF region, $H(\vec{x})=1$ elsewhere.

For all cases we will be using this \textit{hat} constraint, and will define the OFF region as all bins beyond  distance $d$ from the center of the observation. For simplicity's sake we set $d$ as two thirds of an observation's field-of-view radius, but further experiments should attempt varying $d$ or changing OFF region shape altogether to see how results are affected.
Furthermore, to accommodate cases of extended sources where signal is not expected to completely disappear within the field of view, we can instead opt for a \textit{soft hat} that imposes an upper limit (typically a fixed fraction of the background) for the source in the OFF region, rather than making it zero.

\paragraph{Background threshold}
In most realistic cases, source and background are difficult to separate, typically because of overlapping energy spectra, leading to source signal contaminating background estimation, and vice-versa. In extreme cases this can lead to a one-component estimation where source or background is estimated to be zero. Since using the hat constraint prevents background from being zero, a complementary constraint is to impose an upper limit on the background's spatial estimation so that brighter points are assumed to be either due to source signal or Poisson noise; thus limiting the chances of a null source estimation. Since the instrumental background is expected to be spatially isotropic, this is a hypothesis worth using if we know the observation to be relatively spatially homogeneous.
Supposing homogeneous background, we set the mean count per spatial bin as threshold:
\begin{equation}
    T=\big\langle \sum_{\rm ebins}obs \big\rangle_{\rm sbins}
\end{equation}

\begin{figure}
\centering
    \includegraphics[width = \textwidth]{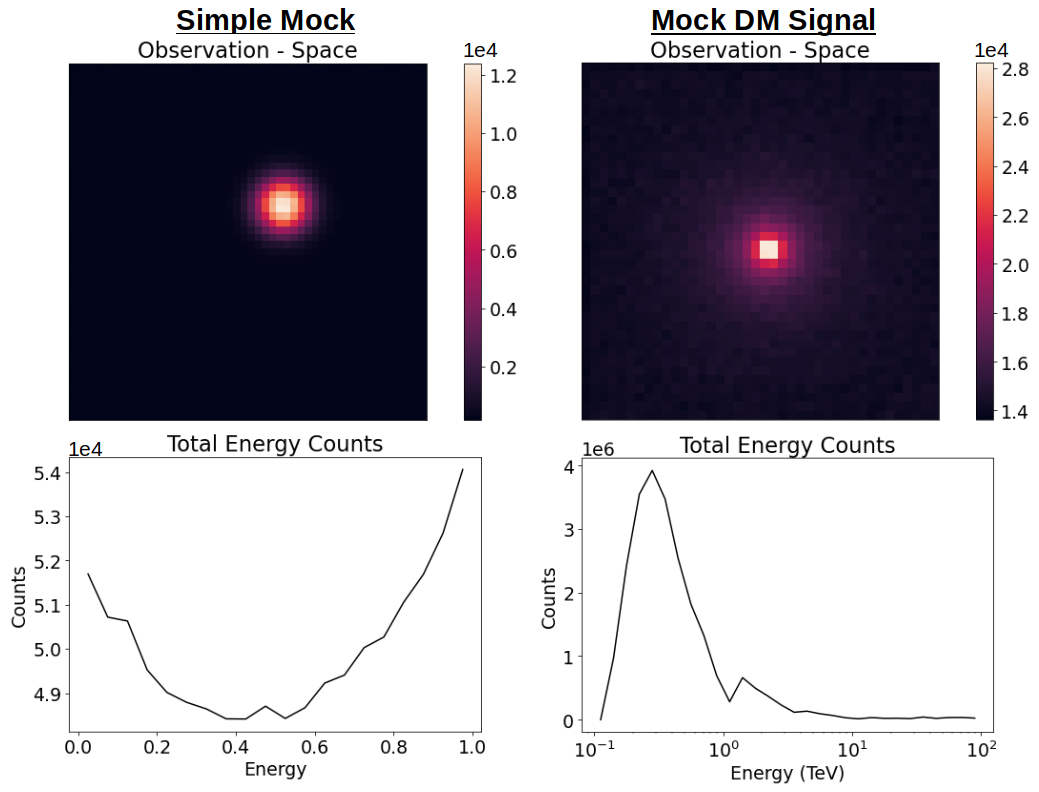}
    \caption{Mock observation for the simple toy model case (left panels) and a simulated observation of dark matter annihilation signals in the Galactic center (right panels). Observations over spatial and energy bins are plotted in the top an bottom panels, respectively. Energy and position are in arbitrary units for the simple mock case.}
    \label{fig:mock_ims}
\end{figure}

\subsection{Rejection Criterion}
As certain cases are bound to be too complex for our model, leading to erroneous estimations, we can perform a simple sanity check to ensure that the final estimation is plausible. Indeed, one need only compare the likelihood of the final total estimated field $F_{\rm tot}^{\rm est}$ to that of the one-component solution $F_{\rm tot}^{\rm 1C}$, which corresponds simply to:
\begin{eqnarray}
    \begin{cases}
       F_{tot}^{1C}=S^{1C}\times E^{1C},\\
       S^{1C}=\sum_{\rm ebins}obs,\\
       E^{1C}=\frac{1}{\sum obs}\sum_{\rm sbins}obs,\\
    \end{cases}
\end{eqnarray}\label{eq:1C}
where the space and energy components simply correspond to the observation summed over energy and space, respectively (normalized for energy).
If $\mathcal{P}(obs|F_{\rm tot}^{\rm 1C})\geq \mathcal{P}(obs|F_{\rm tot}^{\rm est})$, the estimation should be rejected as erroneous.

This being said, in cases with low statistics, the one-component solution as defined above can turn out to have a relatively high likelihood, possibly higher than the ground truth solution, as it overfits the Poisson noise. In such cases comparison against a one-component solution with similar constraints on the smoothness of functions as our two-component models should be preferred.

A more general approach hypothesizing more than one source should similarly consider an iterative likelihood approach wherein the likelihood of an $n$-component solution is compared against that of an optimized $(n-1)$-component solution.

\section{Datasets}\label{sec:dataset}
 
\subsection{Data format}
A typical VHE gamma-ray observation consists of a list of gamma-like events with reconstructed coordinates corresponding to positions in the sky, energy and time.
In what follows we will consider the search for steady emissions and ignore the temporal dimension, but as it is of prime importance in the case of VHE transient physics, it should be considered for subsequent work.
For this study, we work with binned events rather than an event list. Thus our training data, which we henceforth refer to as "observation", corresponds to a 3D histogram with two dimensions in space and one in energy.

\subsection{Mock data}
\subsubsection{Toy model mock data}
We start by building a simple mock observation with optimal conditions. This allows us to establish that the models and method are sound for a simple case and draw comparisons with messier realistic scenarios.

We build a 3D histogram with $50\times 50$ bins in "space" over $[0,1]^2$ and 20 bins in "energy" over $[0,1]$. We set $S_s,S_b,E_s$ and $E_b$ as follows:
\begin{eqnarray}
  \begin{cases}
S_s(\vec{x})=C_s\exp{-(\vec{x}-\vec{x_0})^2},\\
S_b(\vec{x})=C_b,\\
E_s(e)=\alpha_s e^2,\\
E_b(e)=\alpha_b(1-e),\\
  \end{cases}
\end{eqnarray}
where $C_s$ and $C_b$ are constant fixed such that $\int{S_s}=\int{S_b}$, and such that average particle count per bin is 10, and $\alpha_s$ and $\alpha_b$ are normalizing constants such that $\int{E_s}=\int{E_b}=1$ on the considered interval. %

\begin{figure}
\centering
    \includegraphics[width = \textwidth]{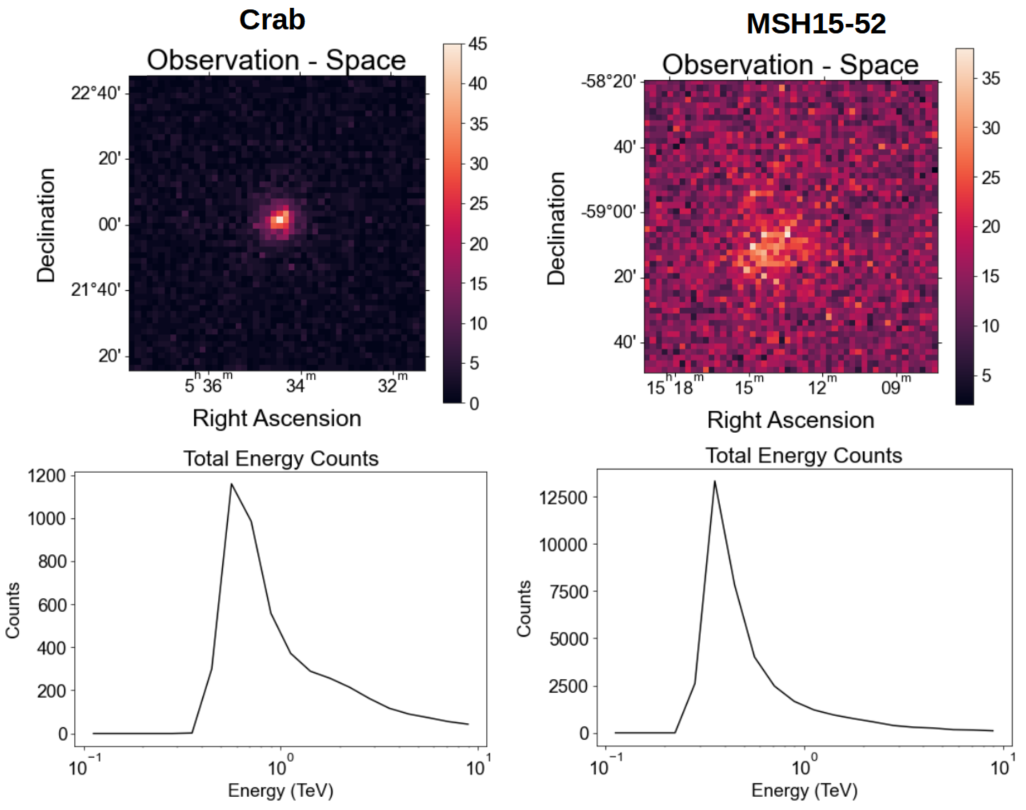}
    \caption{Observations of the Crab nebula (left panels) and the pulsar wind nebula MSH 12-52 (right panels) from the H.E.S.S. public release~\cite{HESSpublicrelease}. Top and bottom panels show the spatial and energy count distributions of the observations, respectively.}
    \label{fig:obs_ims}
\end{figure}

\begin{figure}
\centering
    \includegraphics[width = \textwidth]{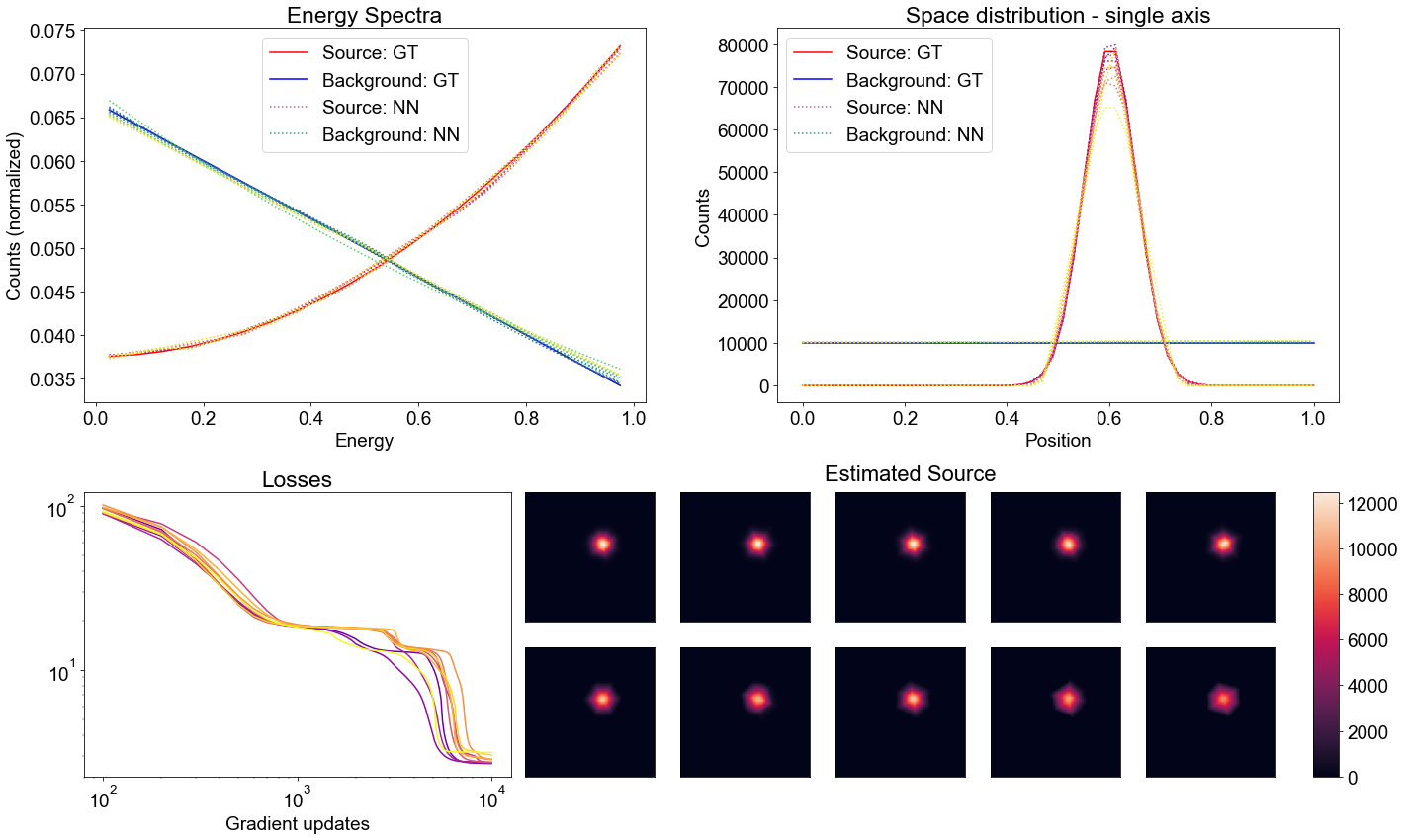}
    \caption{Estimations for the simple mock case for 10 models run with different initial conditions. \textit{Upper left:} Normalized energy counts for source (red) and background(blue). \textit{Upper right:} Total counts for bins over longitude axis. Full lines show ground truth, dotted lines show model results. The lines, ranging from darkest to lightest, represent estimations from highest to lowest likelihood. Energy and position are in arbitrary units. \textit{Bottom left:} Loss evolution for the 10 runs. \textit{Bottom right:} Spatial source estimation for the 10 runs, arranged from highest to lowest likelihood from left to right and top to bottom. }
    \label{fig:simple_mock_all}
\end{figure}

To construct an observation shown in the left-hand side of Fig.~\ref{fig:mock_ims}, we compute a Poisson realization of $F_{tot}(\vec{x},e)$ for each bin $(\vec{x},e)$.
Looking at the energy spectra as shown on the left-hand side of Fig.~\ref{fig:simple_mock_all}, we can note that they are easy to differentiate. The source is also easily identifiable in space, owing to the high source-to-background ($\rm{StB}=\frac{S}{B}$) ratio (up to 60 in the brightest spatial bin) and, consequently, the high signal-to-noise ratio ($\rm{StN}=\frac{S}{\sqrt{B}}$) where the source is present.
All these factors (distinctness of shapes, high StB and StN) are decisive in the ground truth corresponding to the global maximum of the likelihood (given our model constraints) and the models consistently reaching this global maximum.

\subsubsection{Mock dark matter signal in the Galactic Center}
We produce mock data of H.E.S.S. observations of the GC region with the five-telescope array~\cite{HESS:2022ygk}. These observations obtained in the H.E.S.S. Inner Galaxy Survey  provide the highest exposure in the GC region VHE gamma-rays amounting to 546 hours distributed over the inner few degrees of the Milky Way halo. We consider a 2-component mixture of a VHE gamma-ray signal induced by  the annihilation of Majorana dark matter particles populating the DM Milky-Way halo and a residual background component.
The energy-differential residual background flux measured by H.E.S.S. 
is extracted from Ref.~\cite{HESS:2022ygk}, and is assumed to be spatially isotropic. Such an approximation is justified since no significant spatial dependence was found down to 1\% level~\cite{HESS:2022ygk}.  
For more details on the mock data simulation framework, see Ref.~\cite{Montanari:2022buj}.
The signal component will be given by the energy-differential flux of gamma rays in a solid angle $\Delta\Omega$ expected from the pair-annihilation of Majorana DM particles of mass
$m_{\rm MD}$ and in a DM halo of density $\rho$, expressed as:
\begin{equation}\begin{aligned}
\label{eq:dmflux}
\frac{d \Phi}{d E} &=
    \frac{\langle \sigma v \rangle}{8\pi m_{\rm DM}^2}\sum_f  {\rm BR}_f \frac{d N_f}{d E} \, J(\Delta\Omega) \quad {\rm with} \quad
J(\Delta\Omega) &= \int_{\Delta\Omega} d \Omega \int_{\rm los}  d s\, \rho^2(s[r,\theta]) \, .
\end{aligned}\end{equation}

Following the methodology given in Ref.~\cite{Montanari:2022buj}, 
we compute the expected number of signal events for annihilation of DM  of 1 TeV mass in the W$^+$W$^-$ channel with an annihilation cross section of 4$\times$10$^{-25}$ cm$^3$s$^{-1}$,
and a DM density profile following the Einasto parametrization with profile parameters extracted from Ref.~\cite{HESS:2022ygk} in any energy bin $i$ and in any spatial Galactic longitude bin $j$, and spatial Galactic longitude bin $k$. The time exposure map is assumed to have a constant value of 500 hours. The energy resolution is assumed to follow a Gaussian with width given by $\sigma$/E = 10\%. The StB in the brightest spatial bin is equal to 1. The events are binned in 30 logarithmically-spaced bins between 0.1 and 100 TeV (\textit{i.e.} 0.1 dex.), and 40 bins in Galactic longitude in and latitude between $-10^\circ$ and $+10^\circ$.

\subsection{Observational public data}

Finally we aim to test our model on true observational data.
We choose two sources of interest, the Crab Nebula and the pulsar wind nebula MSH~15-52.
We construct our observations by stacking observational runs (i.e. individual observations of the sources over approximately 28 minutes) obtained from public H.E.S.S. data~\cite{HESSpublicrelease} and counting the resulting event list in a 3D histogram.
The Crab and MSH 15-52 datasets correspond to observation times of 1.9 (4 runs) and  9.1 (20 runs) hours, respectively.
We select a field of view of 1.5 degrees centered on the source nominal position and consider events in the energy range from 0.1 to 10 TeV. Within this region, we count events in a histogram made over 50×50 equally spaced spatial bins and 20 logarithhmically-spaced energy bins.
In this configuration, the StB in the brightest spatial bin is estimated to be of around 20 for Crab and 0.8 for MSH 15-52.

\section{Results}\label{sec:results}

\subsection{Simple mock data}
We first show results for the simple mock case (Fig.~\ref{fig:mock_ims}, left). As we can control important parameters such as total event count and signal-to-noise ratio when constructing the data, and know the ground truth distributions for source and background signal, we can accurately determine the performance and limitations of our network.
The model is run ten times with varying randomized initial weights, for $n = 10^4$ gradient updates.
We apply only the \textit{hat} constraint.

In Fig.~\ref{fig:simple_mock_all}, we can note that results are both precise and consistent for all runs, with near-perfect overlap of GT and estimated power spectra (upper left) and space functions (upper right), and loss evolution (lower left) following a similar pattern for every run.

Looking closely at the spatial source estimations (bottom right), we can note the networks' limitation in approximating a 2D gaussian, exhibiting more of a branched star-like pattern; this is to be expected given our networks' relatively simple architecture. However we will see in the following that averaging over a sufficient number of runs mitigates this effect.

\begin{figure}
\centering
    \includegraphics[width = \textwidth]{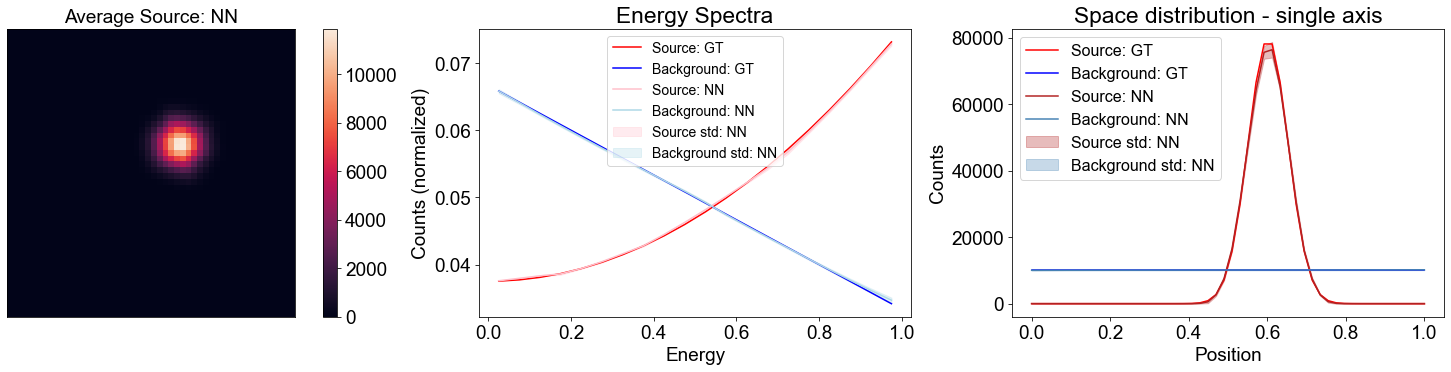}
    \caption{Weighted average estimations for the simple mock case for hat constraint. Left: average source space estimation. Middle: averaged energy functions with standard deviation. Right: averaged spatial functions over longitude axis with model standard deviation. Energy and position are in arbitrary units.}
    \label{fig:simple_mock_avg}
\end{figure}

We compute a weighted average of the recovered space and energy fields for the ten runs. 
For all fields $F$ (corresponding to $S_s,S_b,E_s$ and $E_b$), $F_{avg}$ is defined as $F_{\rm avg}=\sum_{\rm i} w_{\rm i} F_{\rm i}$ where $w_{\rm i}$ is the weight defined in Eq.~(\ref{eq:ponderation}). Correspondingly for all $F$ we compute a weighted uncertainty $\nu_F=\sum_{\rm i} w_{\rm i} |F_{\rm avg}-F_{\rm i}|$, which can be likened to our model's epistemic uncertainty. 
Looking at the recovered average fields from Fig.~\ref{fig:simple_mock_avg}, we find that the space gaussian is more closely approximated, while estimations both in space and energy show near perfect overlap with ground truth, with negligible model uncertainty. 

Defining the weighted relative difference in the following manner:
\begin{equation}
    D_R=\frac{\int |F_{\rm{NN}}-F_{\rm{GT}}|}{\int |F_{\rm{GT}}| },
\end{equation}\label{eq:rdiff}

the models recover the energy spectra (center) within $0.4\%$ for source and $0.3\%$ for background, and the space functions (left/right) within $6\%$ for source and $1.4\%$ for background.

\subsection{H.E.S.S. public data}
We now compute results for real observational data from the H.E.S.S. public release.
Since we do not have a ground truth for comparison, we compare our field estimations with those obtained using standard methods (SM); for energy spectra estimation this corresponds to the MultipleOFF method \citep{Berge:2006ae}, whereas for spatial estimation we use the ring background method \citep{Berge:2006ae}.

\subsubsection{The point-like Crab nebula}
\begin{figure}[h]
\centering
    \includegraphics[width = \textwidth]{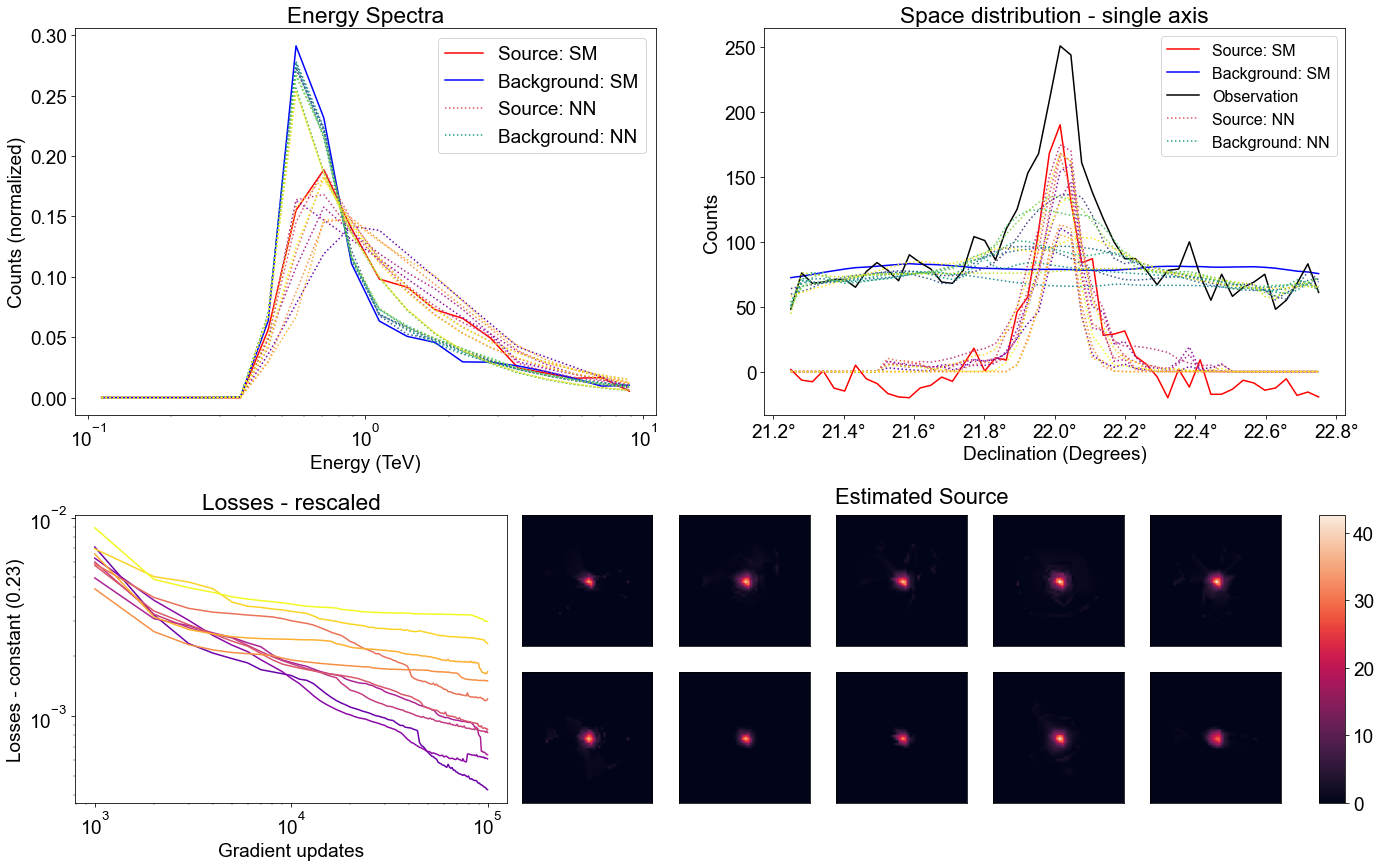}
    \caption{Estimations for the Crab nebula for 10 models run with different initial conditions. \textit{Upper left:} Normalized energy counts in log-bins for source (red) and background (blue). \textit{Upper right:} Total counts for bins over longitude axis. Full lines show standard model results, dotted lines show NN model results. The lines, ranging from darkest to lightest, represent estimations from highest to lowest likelihood. \textit{Bottom left:} Loss evolution for the 10 runs. \textit{Bottom right:} Source space estimation for the 10 runs, arranged from highest to lowest likelihood from left to right and top to bottom.}
    \label{fig:crab_hat_all}
\end{figure}

We first observe results for 10 individual runs as shown in Fig.~\ref{fig:crab_hat_all}, using only the \textit{hat} constraint. 
Looking at the energy spectra (upper-left panel), we can see that estimations for background spectra show little variation and are consistent both with one another and with SM.
This is to be expected given our hat constraint, which defines a background-only OFF region, thereby constraining the background spectrum. Comparatively, we can see more variation in estimations of the source spectrum; however they remain consistent with SM measurements.

Looking at spatial source estimations (bottom right panel), results seem somewhat consistent with one another, recovering a Gaussian of slightly varying amplitude and width, along with surrounding low-amplitude excess.
A one-dimensional spatial observation of source and background (upper right panel) reveals the reason for the varying amplitude of the estimated source, as we can see the source signal more or less "contaminating" the estimated background, decreasing the corresponding source estimation in the process. This can be mitigated by thresholding the background.
The spatial NN estimations are somewhat consistent with SM, 
though comparison suggests that NN estimations more closely fit total observed signal, while still smoothing out Poisson fluctuations.

The loss functions (bottom left panel) are rescaled ($L'=L-c$) to better show their spread (here $c=0.23$). There is very little difference between final losses, and therefore between the likelihoods of the final estimations.

\begin{figure}
\centering
    \includegraphics[width = \textwidth]{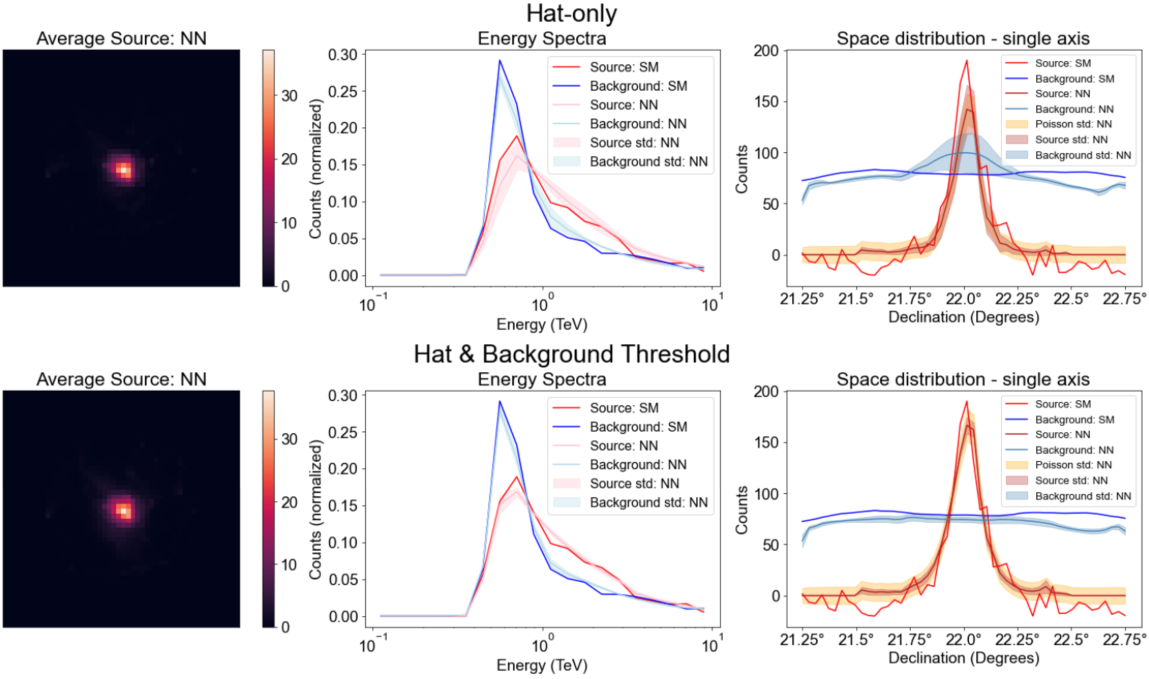}
    \caption{Weighted average estimations for the Crab Nebula for the 
    \textit{hat}-only (top panels) and  \textit{hat + bck threshold} (bottom panels) constraints, respectively. 
    \textit{Left panel:} average source space estimation. \textit{Middle panel:} averaged energy functions with standard deviation. \textit{Right panel:} averaged spatial functions over longitude axis with Poisson and model standard deviation.}
    \label{fig:crab_hat_avg}
\end{figure}

As before we observe the likelihood-weighted average for all fields as shown in  Fig.~\ref{fig:crab_hat_avg}. Looking first at the \textit{hat}-only constraint case (top), we can see that the average spatial source estimation (left panel) is smooth in shape, with the low-amplitude excesses of individual estimations mostly vanishing.
Looking over one spatial dimension (right panel), we can see that the average estimated background signal remains apparently contaminated by source signal, although a flat estimation is within our method's region of uncertainty (as defined by weighted standard deviation). This in turn reduces average estimated source signal, creating a visible inconsistency with SM.
Looking at the averaged energy spectra (center panel) we can see that estimations are consistent with SM, with a larger uncertainty and disparity with SM for the source.

Upon adding a threshold constraint on the background (bottom panel), we find that uncertainty is largely decreased for all estimations, with near perfect agreement with SM for the energy spectra (center panel), and in space (right), SM's estimated excess mostly falling within one Poisson deviation (shown as a yellow-shaded region) of the NN's average estimated source. This improved agreement makes sense, given that we are applying similar constraints to that of the SM.

\subsubsection{The extended source MSH 15-52}

\begin{figure}[h]
\centering
    \includegraphics[width = \textwidth]{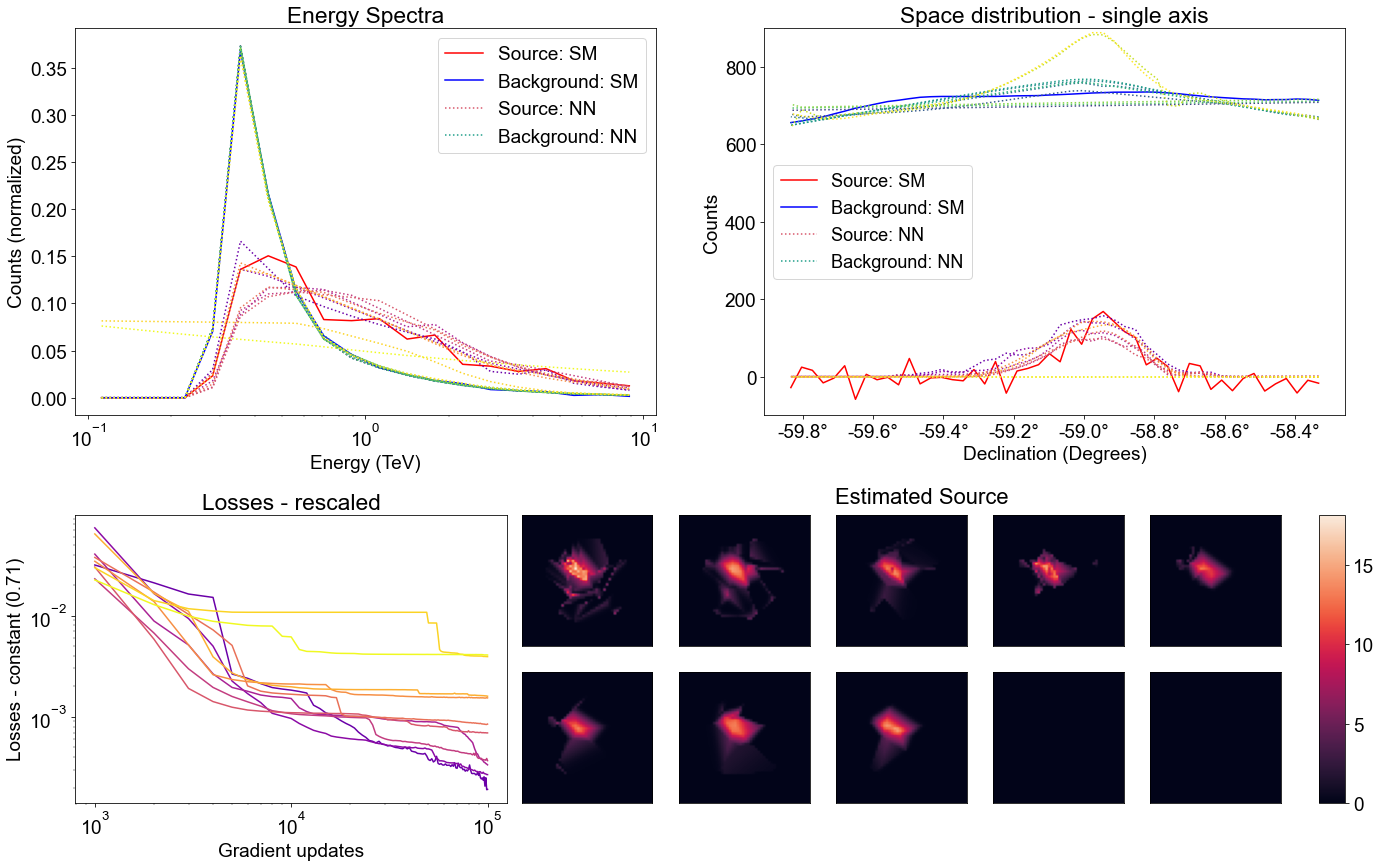}
    \caption{Estimations for the pulsar wind nebula MSH15-52 for 10 models run with different initial conditions. \textit{Upper left:} Normalized energy counts in log-bins for source (red) and background (blue). \textit{Upper right:} Total counts for bins over longitude axis. Full lines show standard model results, dotted lines show NN model results. The lines, ranging from darkest to lightest, represent estimations from highest to lowest likelihood. \textit{Bottom left:} Loss evolution for the 10 runs. \textit{Bottom right:} Source space estimation for the 10 runs, arranged from highest to lowest likelihood from left to right and top to bottom.}
    \label{fig:MSH15_hat_all}
\end{figure}

We first observe results for 10 individual runs as shown in Fig.~\ref{fig:MSH15_hat_all}, using only the \textit{hat} constraint. 
For the energy spectra (top-left panel), estimations for background spectra barely show any variation and perfectly overlap with SM estimation. This is due to the long observation time which provides increased statistics, allowing for a precise estimation of the background spectrum.

Here again, while still consistent with SM, the source estimations show more variation. We can note two spurious cases (yellow and light orange curves), inconsistent with the other estimations. These correspond to the cases with highest final losses (lower left panel), i.e. with lowest final likelihoods. Looking at the corresponding spatial estimations, it becomes clear that these correspond to a one-component solution; the estimated source in space is zero (bottom right panel, bottom right images), and looking at the one-dimensional case (upper right panel) we can see that the equivalent background estimations correspond to the total field.
\begin{figure}
\centering
    \includegraphics[width = \textwidth]{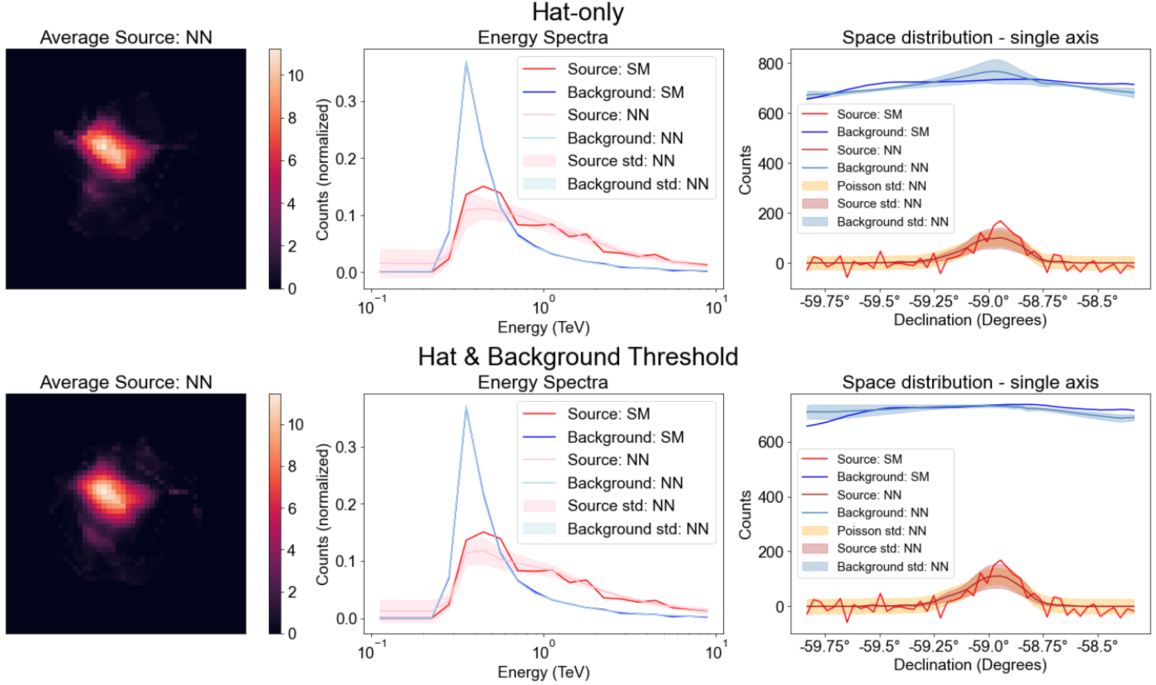}
    \caption{Weighted average estimations for the MSH~15-52 for the \textit{hat}-only (top panels) and  \textit{hat + background threshold} (bottom panels) constraints, respectively.
     \textit{Left panel:} average source space estimation. \textit{Middle panel:} averaged energy functions with standard deviation. Right: averaged space functions over longitude axis with Poisson and model standard deviation.}
    \label{fig:MSH1552_avg}
\end{figure}

We now observe the likelihood-weighted average for all fields as shown in Fig.~\ref{fig:MSH1552_avg}.
In the \textit{hat}-only constraint case (top panels), the average spatial source estimation (left panel) is somewhat smooth in shape, with some low-amplitude excess remaining. More runs might be warranted to smooth these out.
Looking over one spatial dimension (right panel), the average estimated background signal is once more contaminated by source signal, especially due to the previously mentioned one-component estimations affecting the average. A flat estimation is within our method's region of uncertainty as defined by weighted standard deviation. As in the Crab case this reduces average estimated source signal, creating a slight inconsistency with SM. Still, SM-estimated excess is essentially within a Poisson deviation of NN-estimated source as shown by the yellow field.
In the averaged energy spectra (center panel) we can see that background estimation overlaps perfectly with SM, while source estimation shows more discrepancy with SM due to one-component estimations. 

Putting a threshold on the background (bottom panel) forces a two-component solution, removing the issues brought up in the previous case; the estimated background is flatter, all NN estimations are more consistent with SM. However unlike for the Crab case, source estimations maintain a high uncertainty even with this additional constraint, likely due to the complex shape and fainter emission of the source.

\subsection{Mock diffuse dark matter signal}
\begin{figure}[htbp]
    \centering
    \includegraphics[width=\textwidth]{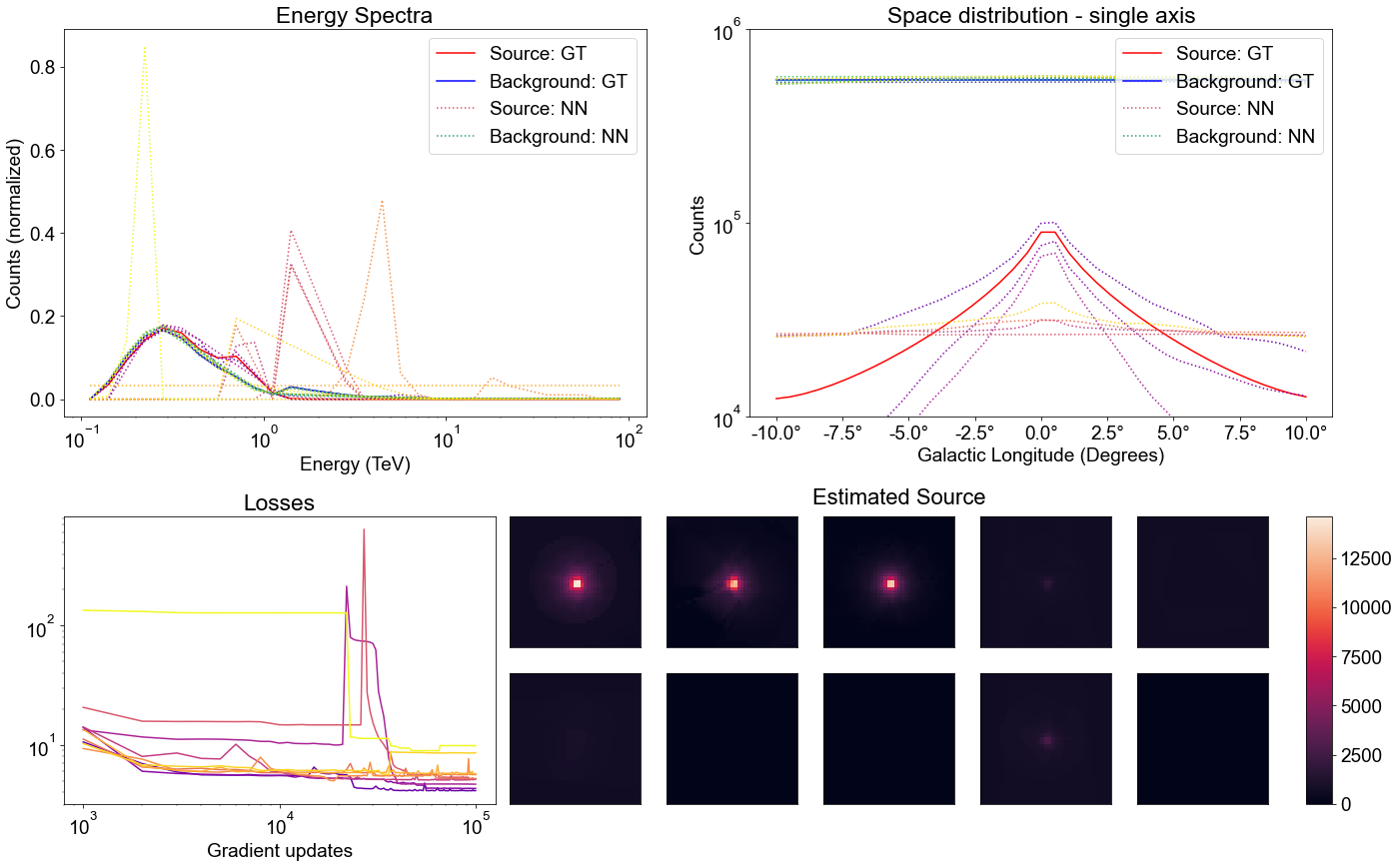}
    \caption{Estimations for the DM Mock data for 10 models run with different initial conditions. \textit{Upper left:} Normalized energy counts in log-bins for source (red) and background (blue). \textit{Upper middle:} Total counts for bins over longitude axis. Full lines show ground truth, dotted lines show model results. The lines, ranging from darkest to lightest, represent estimations from highest to lowest likelihood. \textit{Bottom left:} Loss evolution for the 10 runs. \textit{Bottom right:} Source space estimation for the 10 runs, arranged from highest to lowest likelihood from left to right and top to bottom.}
    \label{fig:DM_mock_all}
\end{figure}
We examine here results for the more complex case of the mock DM signal. Given the generally low signal-to-background ratio, it is necessary to use both hat and background threshold constraints to obtain satisfactory results. Since the source is extended such that it covers the entire field of view, we opt for a "soft" hat, where spatial source signal is thresholded to $5\%$ of spatial background in the OFF region.

We first observe results for 10 individual runs as shown in Fig.~\ref{fig:DM_mock_all}.
For the energy spectra (top-left panel), estimations for background spectra show once again little variation and overlap with GT, due to the \textit{hat} constraint.
On the other hand, source estimations show significantly more variation, with all but the highest-probability estimations showing significant discrepancy with GT. These aberrations can be explained by the relatively low total source signal, which implies low impact on total likelihood estimation even in very erroneous cases.
Inspecting the spatial estimations (top and bottom right panels), we can note that background estimations are generally identical and consistent with GT. Conversely, source estimations can be divided into two categories; essentially flat estimations where source value corresponds to the imposed hat threshold, and pointed estimations that better capture GT signal shape. We can note that the second category corresponds to higher likelihood scores, which will be relevant when computing a weighted average.
Observing the losses (bottom left panel), we can see that most first-category (lower-likelihood) cases stagnate throughout training; this is likely due to the early highly erroneous estimations of energy spectra preventing a gradual convergence of spatial estimations towards GT. These cases correspond to local minima.

\begin{figure}[h]
    \centering
    \includegraphics[width=\textwidth]{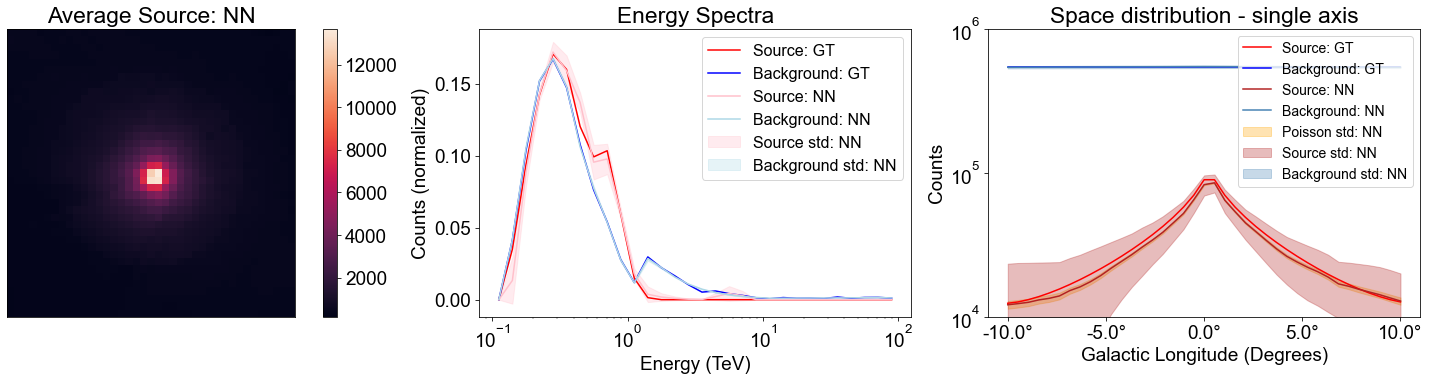}
    \caption{Weighted average estimations for the DM Mock data, for hat + background threshold constraints. Left: average source space estimation. Center: averaged energy functions with standard deviation. Right: averaged space functions over longitude axis with model (red) standard deviation.}
    \label{fig:DM_mock_avg}
\end{figure}

Looking at the likelihood-weighted average for all fields shown in Fig.~\ref{fig:DM_mock_avg}, the importance of weighting takes its full measure, as the averaged estimated source fields suddenly become near-perfectly consistent with GT.
Indeed, we can see that the average spatial source estimation (left panel) is smooth in shape, that energy functions (center panel) overlap closely (with an average relative difference of $7.6\%$ for source and $1.6\%$ for background) and with little variation for both source and background, and that 1D space estimations (right panel), while showing more significant uncertainty for source, overlap very well on average with GT, with a relative difference in 2D space of $10\%$ for source and $0.4\%$ for background.
A significant finding is that the model's prediction of the spectral shape of the signal component accurately captures the bump-like gamma-ray feature of the $W^+W^-$ annihilation channel near the DM mass. This advancement allows for the reconstruction of DM-induced features in a fully non-parametric manner. It would benefit DM searches, particularly within realistic DM models where multiple annihilation channels, weighted by their branching ratios, contribute to the final-state gamma-ray spectrum, in particular at energies close to the DM mass.

Finally, inspecting the weights (\textit{i.e.} likelihoods) of individual estimations, we find that only the top three cases contribute significantly to the weighted average, while more abhorrent estimations are conferred negligible weight, allowing for excellent results despite many "failed" runs.
This in turn brings up the necessity of doing a sufficient amount of runs. Indeed, the more difficult a problem (low StB, low statistics, overlapping spectra, lighter constraints) the more runs it will take to find and better cover global maxima for the likelihood. 

\section{Discussion}\label{sec:conclusion}

In this paper we developed a NN-based deep ensemble approach for source-background separation in noisy very-high-energy gamma-ray observations.
This method consists in training $n$ sets of networks to estimate the spatial and energy components for two fields corresponding to the average source and background, so as to maximize the likelihood for the observation to be a realization of their sum (the total estimated field).
Constraints are imposed on the estimations to better differentiate source and background, as well as to render the task easier. One such constraint (\textit{hat}) consists in defining and imposing an OFF region in space where the source is made to be negligible, and another (\textit{bkg threshold}) consists in imposing a threshold on the spatial background intensity, such that brighter regions have to be considered as due to source or Poisson fluctuation of the background by the NN.
The $n$ runs are then weighted according to their likelihood and averaged to obtain a final estimation for all components, as well as an uncertainty corresponding to their weighted standard deviation.
We tested this approach both on simple and more complex realistic mock observations, as well as real observations from the H.E.S.S. public data release. We compared results to ground truth in mock cases and results derived from standard methods in real cases.
We found that our approach worked remarkably well in all considered cases, with average component estimations showing good to near-perfect agreement with ground truth and standard method, and with performance clearly improving when combining constraints.

Going forward, it is important to keep in mind a few limitations in our approach.
For one, the model relies on strong hypotheses: that there are only two signals in the observation, that space and energy components in the signals are separable, and, depending on the constraints we choose, that we can confidently define an OFF region in an observation, and that the observation is "flat" enough (i.e. relatively homogeneous acceptance rate with respect to space in considered observation) that we can posit a threshold for the background.
These assumptions can reasonably be made for all the cases considered in this article, however in larger grid-surveys, we can expect more inhomogeneity in observations to arise due to varying conditions such as weather, changing azimuthal angles and varying energy acceptance with respect to observation angle causing space/energy mixing. We can also expect more than one source per observation, prompting the need for a more generalized structure that could allow for multiple sources. 
Additionally, all cases considered have relatively high ($>0.1$) StB throughout space, with sufficient exposition that StN is $>1$ throughout the observation's considered space.
More difficult cases, such as dividing source signal strength by ten for the mock DM data, did not provide satisfactory results within similar training time frames and number of runs.
Indeed, in such cases, we can expect models to be more likely to fall into local maxima, while reaching global maxima could take more time, requiring longer training and more attempts. As this is time- and energy-consuming, methods to optimize training and better tackle the issue of local minima (adapting training rate, simulated annealing~\cite{bertsimas1993simulated}, detecting stagnation to stop training early, sequential training, etc) should be considered before addressing such cases.

Further work on this subject could take several paths.
Most obvious would include steps towards generalization, such as adding a temporal dimension, or allowing for a varying number of sources.
While adding the option for additional sources should be fairly easy to implement in code (typically by ensembling runs that alternatively hypothesize $1,2,..,n$ sources, and looking at maximally likely estimations), this comes with new challenges. Namely, if no discriminating constraints are applied for different sources, this may lead to several cases of degeneracy (\textit{e.g.} one single source being interpreted as a sum of two, or network 1 alternatively estimating source A or B and conversely for network 2, posing an issue when averaging). This is to say nothing of the added leeway to overfit Poisson noise due to the additional degrees of freedom, and, in parallel, the increased difficulty to find the global maxima for the problem. While the multiple source problem is an interesting and necessary path to explore, one must tread carefully, making sure to have good criteria when comparing likelihoods for estimations with different degrees of freedom, and using discriminating constraints whenever possible.
Conversely, another path would be to further make use of physical knowledge for more precise estimations. This could include taking into account additional instrumental input like space and energy uncertainty in events, which could help to smooth out Poisson noise, or information such as instrument response function. Alternatively this approach could be used in tandem with other analysis methods to better define constraints or a-priori hypotheses like number of sources or assumed OFF regions.
Finally, experimenting with various architectures, within a set of deep ensemble runs or for all sets, should be instrumental in optimizing results. Dense NNs are interesting in the context of a proof of concept, but one could expect a convolutional neural network (CNN, \cite{o2015introduction}) taking the observation as input as well as a base of comparison with the output to compute the loss (as in Ref.~\cite{PhysRevLett.125.241102}), our model only doing the latter, to be more suited to the task. 
One might also explore more direct non-parametric methods for fitting the observation histogram, such as using wavelets with smoothness constraints \citep{rioul1991wavelets}.

When aiming to solve the problem of source-background separation in noisy observations, we sought a method to find and display a solution, or rather a manifold of solutions, that maximizes the likelihood of the observation while optimally exploring function space and displaying model uncertainty, as could be expected in noisy observations.
We found that our approach, using a deep-ensemble set of low-parameter dense nets, proved greatly effective in this regard. Critical to obtaining optimal and accurate final estimations were two key factors.
The first was the use of informative constraints on source and background shapes. These constraints helped guide the model towards more realistic and reliable solutions.
The second, and a key finding of this work, was the use of an adequate weighting method when combining individual results of the deep ensemble. This method allowed us to automatically discard more erroneous estimations while enabling similarly plausible estimations to contribute to the final average.
In conclusion, our methodology demonstrates a robust framework for source-background separation in the presence of noise, leveraging the strengths of deep ensembles and effective weighting techniques. This approach not only enhances the accuracy of the results but also provides a clear representation of model uncertainty, paving the way for more reliable and insightful analyses in noisy observational data. Future work can build on these findings to further refine and expand the applicability of our approach to a broader range of astrophysical problems.

\acknowledgments

The authors would like to thank Alexander Belikov and Alessandro Montanari, as well as Mathieu de Bony de Lavergne for interesting and informative conversations, and Jérôme Bobin for his helpful advice. We are also grateful to the anonymous referee for their valuable and insightful comments, which have significantly improved this manuscript.
This project was funded by the \textit{Programme Transversal de Compétences Simulation Numérique} of CEA.

\bibliographystyle{JHEP}
\bibliography{references}

\end{document}